\documentclass[aps,prd,floatfix,groupedaddress,nofootinbib,superscriptaddress,twocolumn,notitlepage,showkeys]{revtex4-2}

\usepackage{amsmath,amssymb,bm,color,float,graphicx,mathrsfs}
\usepackage[normalem]{ulem}
\usepackage[hidelinks]{hyperref}
\hypersetup{colorlinks=true,linkcolor=blue,citecolor=blue,urlcolor=blue}
\usepackage{Macro}
\usepackage{CJKutf8}
\usepackage{orcidlink}

\def\MB{{\cal B}}
\def\MH{{\cal H}}
\def\bq{\bm{q}}

\begin{document}

\title{Cross-correlation between CMB B-modes and Faraday Rotation as a Novel Probe of Primordial Magnetic Fields}

\author{Toshiya Namikawa}
\affiliation{Kavli IPMU (WPI), UTIAS, The University of Tokyo, Kashiwa, 277-8583, Japan}

\author{\begin{CJK*}{UTF8}{gbsn}Guanming Liang (梁冠铭)\orcidlink{0000-0002-2581-2406}\end{CJK*}}
\affiliation{Department of Applied Mathematics and Theoretical Physics, University of Cambridge, Wilberforce Road, Cambridge CB3 0WA, United Kingdom}
\affiliation{Kavli Institute for Cosmology, University of Cambridge, Madingley Road, Cambridge CB3 0HA, United Kingdom}

\date{\today}

\begin{abstract}
We investigate the cross-bispectrum between the cosmic microwave background (CMB) $B$-mode polarization and two Faraday rotation (FR) angles as a new probe of primordial magnetic fields (PMFs). Assuming Gaussian PMFs with both non-helical and helical components, we derive the $B\alpha\alpha$ bispectrum and decompose it into parity-even and parity-odd contributions. The mixed non-helical--helical contribution appears in the parity-even sector, whereas the purely non-helical and purely helical contributions contribute to the parity-odd sector. This parity dependence provides sensitivity to PMF helicity that is absent in the FR power spectrum alone. We then forecast the detectability of the cross-bispectrum for SO-like and LiteBIRD-like CMB experiments, and find that future observations can probe PMFs with amplitudes at the sub-nG level. The $B\alpha\alpha$ bispectrum is dominated by the helicity-sensitive parity-even sector and provides a promising avenue for constraining both the amplitude and helicity of PMFs.
\end{abstract} 

\keywords{cosmology, cosmic microwave background}


\maketitle

\section{Introduction} \label{sec:intro}

The origin of the $\mu$G magnetic fields observed in galaxy clusters remains incompletely understood \cite{Widrow:2002ud,Durrer:2013pga}. One of the possible origins is the so-called primordial magnetic fields (PMFs) that are generated in the early universe during inflation or phase transitions (see \cite{Kandus:2010:review,Vachaspati:2020blt} and references therein). Multiple studies have argued that observations of gamma rays from distant sources can also be interpreted as an indirect evidence for PMFs coherent over Mpc scales in cosmic voids \cite{Neronov:2010,Tavecchio:2011,Vovk:2012,Takahashi:2013lba,Veres:2017aou}, the existence of which is difficult to explain via galactic dynamos alone. A nonzero helicity of void magnetic fields has been suggested based on the gamma-ray arrival directions observed by the Fermi Large Area Telescope \cite{Tashiro:2013:helical-PMF,Chen:2014qva,Tashiro:2014gfa}. 
More critically, there must be PMF helicity to initiate the ``inverse cascade" mechanism \cite{Banerjee:2004df,Campanelli:2007tc}, without which magnetohydrodynamic simulations show that turbulent dissipation causes the comoving PMF amplitude to decay far more rapidly than the standard inverse scale-factor squared dilution of cosmic expansion (making PMFs too weak to contribute non-trivially to the cosmological magnetic fields observed). Furthermore, it has been postulated that the Hubble tension can originate from not accounting for PMFs in recombination physics \cite{Jedamzik:2020krr, Jedamzik:2023csc,Jedamzik:2025cax}. These observations motivate the possibility that at least part of the observed cosmic magnetic fields originated from PMFs generated in the early Universe and further observational tests for PMFs are important.

Measurements of the anisotropies of the cosmic microwave background (CMB) have played a key role in constraining PMFs. PMFs contribute to the total energy momentum tensor of the cosmological plasma, gravitationally modify the CMB anisotropies, and thus, the CMB angular power spectra \cite{Subramanian:1998fn,Durrer:1998ya,Durrer:1999:PMF-tensor,Mack:2001gc,Subramanian:2002nh,Subramanian:2003sh,Lewis:2004,Giovannini:2004aw,Yamazaki:2004vq,Kahniashvili:2006:CMB-PMF,Giovannini:2007sr,Giovannini:2007qn,Giovannini:2008yz,Yamazaki:2008:CMB-PMF,Finelli:2008:CMB-PMF,Paoletti:2008,Shaw2010MassiveUniverse,Bonvin_2010,Kunze:2011:CMB-PMF,Paoletti:2011,Shaw&Lewis:2012,Paoletti:2013,Zucca:2016iur}. 
The measured CMB angular power spectra by Planck, BICEP/Keck Array, and the South Pole Telescope placed an upper bound on the strength of the nearly scale-invariant magnetic field smoothed over a region of $1$\,Mpc comoving scale as $B_{1\rm Mpc}\leq 1.5^{+0.9}_{-1.4}$\,nG (2\,$\sigma$) \cite{Paoletti:2019pdi}.
The PMF-induced non-Gaussianity in the CMB primary anisotropies leads to a comparable constraint on $B_{1\rm Mpc}$ \cite{Seshadri:2009,Caprini:2009:non-Gauss,Trivedi:2011vt,Shiraishi:2011fi,Shiraishi:2011dh,Shiraishi:2012rm,Shiraishi:2013wua,Trivedi:2013wqa,Planck:2015zrl}. The effects of helical PMFs on the CMB primary anisotropies have been also studied through the CMB angular power spectra~\cite{Caprini:2003vc,Kahniashvili:2005xe,Kunze:2011,Ballardini:2014jta}, bispectra~\cite{Shiraishi:2012:helical-PMF}, and trispectra~\cite{Yura:2025}. Constraints on helical PMFs have also been obtained from the Wilkinson Microwave Anisotropy Probe (WMAP) and Planck CMB power spectra \cite{Kahniashvili:2014:Herical,Planck:2015zrl} 

Faraday rotation (FR) of CMB polarization provides another sensitive probe of PMFs in CMB experiments. FR rotates the CMB polarization angle and converts part of the primary CMB $E$-modes into $B$-modes. FR has been studied as a probe of PMFs in Refs.~\cite{Pogosian:2001,Ensslin:2003ez,Campanelli:2004:FR,Kosowsky:2004zh,Planck:2015zrl}, and the FR-induced polarization power spectra measured from WMAP at low frequencies were used to constrain the PMFs with an uncertainty of $\mC{O}(100)$\,nG \cite{Kahniashvili:2008:FR,Pogosian:2011qv}. 
A more powerful way to constrain FR is to directly reconstruct FR angles using correlations of the $E$ and $B$ modes at different angular scales \cite{Kamionkowski:2008fp,Pogosian:2013dya,Namikawa:2016:rot}. The angular power spectrum of the reconstructed FR angles is a four point correlation of CMB polarization fields \cite{Namikawa:2016:rot} and has been used to constrain PMFs \cite{POLARBEAR:2015:rot,BICEP2:2017lpa}. The recent polarization data by BICEP/Keck Array placed the constraint on PMFs through the direct reconstruction of the FR power spectrum as $B_{1\,{\rm Mpc}}\leq 6.6\,$nG ($2\,\sigma$) at $95\,$GHz \cite{BICEPKeck:2022:LoS-dist}. The FR power spectrum is, however, insensitive to the helical component of PMFs. Nevertheless, measurements of CMB FR are expected to become increasingly sensitive to PMFs with future CMB observations \cite{Pogosian:2013dya,Mandal:2022tqu}.
The measurements of FR will constrain $B_{1\rm Mpc}$ with the uncertainty of $\mC{O}(0.1)$\,nG in the Simons Observatory (SO) and LiteBIRD \cite{Pogosian:2019jbt,Mandal:2022tqu}, and of $\mC{O}(0.01)$\,nG in the CMB-HD experiment \cite{Mandal:2022tqu,CMB-HD:2022bsz}. 

The correlation between the primary CMB anisotropies and FR, however, has not yet been explored. Such cross-correlations offer several advantages over measurements based on either observable alone. First, they have different sensitivities to instrumental and astrophysical systematics and therefore provide an independent consistency check of PMF constraints obtained from the primary CMB anisotropies or FR. For example, FR from the Galactic magnetic fields could be a potential systematic in the measurements of PMF-induced FR in future CMB experiments \cite{Oppermann_2012,De:2013dra}. The $B$-mode polarization could be dominated by the Galactic foregrounds. The cross-bispectrum at least should have different response to the Galactic contributions and is useful for a cross-check of the PMF constraint. 
More importantly, although the FR power spectrum itself is insensitive to magnetic helicity, its correlation can retain information about the helical component of PMFs. Combining the tensor-FR-FR bispectrum with the B-mode or FR power spectrum thus breaks the degeneracy between helical and non-helical PMF contributions. Cross-correlations between the primary CMB and FR may therefore provide a promising new avenue for probing the helical part of PMFs.

In this paper, we investigate the bispectrum constituting one CMB $B$-mode polarization and two FR angles as a new probe of PMFs. We derive the bispectrum induced by PMFs, including their helical component, and forecast its signal-to-noise ratio for future CMB experiments. We show that the parity-even bispectrum is particularly useful for isolating the contribution linear in magnetic helicity. 

The remainder of this paper is organized as follows. Section~\ref{sec:CMB-PMF} reviews the generation of the primary CMB anisotropies and FR by PMFs. In Sec.~\ref{sec:bispec}, we derive the cross-bispectrum involving the CMB $B$-mode polarization and FR. Section~\ref{sec:forecast} presents forecasts for the signal-to-noise ratio and discusses the resulting sensitivity to PMFs. We summarize our results and conclude in Sec.~\ref{sec:summary}.

Throughout this paper, we make frequent use of the vector and tensor spherical harmonics $Y_{\ell m,i}^{L}$, $Y_{\ell m,i}^{E}$, $Y_{\ell m,i}^{B}$, $Y_{\ell m,ij}^{TB}$, and $Y_{\ell m,ij}^{VB}$~\cite{Tomita:1982,Kosowsky:2004zh,Dai:2012bc}. Our conventions and useful identities are summarized in Appendix~\ref{app:harmonics}.

\section{CMB anisotropies and Faraday rotation from PMFs} 
\label{sec:CMB-PMF}

We summarize how the $B$-mode polarization and Faraday rotation are related to the magnetic fields. 

\subsection{B-mode polarization}

In general, CMB B-mode polarization from tensor perturbations is given by \cite{Shiraishi:2010:alm-expression}
\al{
    B_{\l m} = 4\pi(-\iu)^\l\sum_{\lambda=\pm2}\frac{\lambda}{2}\Int{3}{\bk}{(2\pi)^3}\Delta_\l^{B,T}(k)h^{(\lambda)}(\bk)Y_{\l m}^{-\lambda,*}(\hk)
    \,, \label{Eq:def-B-mode}
}
where $h^{(\pm2)}(\bk)$ is the tensor metric perturbations, $Y_{\l m}^{\pm 2}(\hk)$ is the spin-2 spherical harmonics and $\Delta_\l^{B,T}(k)$ is the B-mode transfer function for tensor. 
The tensor metric perturbation from the PMF tensor passive mode is sourced by the primordial PMF curvature perturbation contracted with the transverse-traceless polarization tensor $e_{ij}^{(-\lambda)}$ \cite{Shaw2010MassiveUniverse,Shiraishi:2012rm}
\al{
    h^{(\lambda)}(\bk) 
    &= -R_\gamma\left(\ln\frac{\eta_\nu}{\eta_B}\right)\frac{3}{4\pi\rho_{\gamma,0}}
    \notag \\
    &\quad\times e_{ij}^{(-\lambda)}(\hk)\Int{3}{\bq}{(2\pi)^3}\MB^i(\bm{q})\MB^j(\bk-\bq)
    \,, \label{Eq:h-PMF}
}
where $R_\gamma$ is the ratio of the energy density of photons to all relativistic particles, $\eta_\nu$ and $\eta_B$ are the conformal time at the neutrino decoupling and the PMF production, respectively, $\rho_{\gamma,0}$ is the current energy density of photons, and $\MB^i(\bk)=\MB^i(\eta,\bk)a^2$ is the Fourier mode of the comoving magnetic fields.  
Using the unit polarization vector $\epsilon^{\pm}_i(\hk)$ defined in Appendix \ref{app:harmonics}, we define 
\al{
    e^{(\pm2)}_{ij}(\hk) = \sqrt{2}\epsilon^{\pm}_i(\hk)\epsilon^{\pm}_j(\hk)
    \,. 
}
This satisfies the orthogonal relation; $e_{ij}^{(+2)}(\hk)e^{(-2),ij}(\hk)=e_{ij}^{(-2)}(\hk)e^{(+2),ij}(\hk)=2$. 
Substituting Eq.~\eqref{Eq:h-PMF} into Eq.~\eqref{Eq:def-B-mode}, we obtain
\al{
    B_{\l m} 
    &= \Int{}{k}{} T_\l^{B}(k) \Int{2}{\hk}{(2\pi)^3} Y^{TB,*}_{\l m,ij}(\hk) 
    \notag \\
    &\quad \times \Int{3}{\bq}{(2\pi)^3}\MB^i(\bq)\MB^j(\bk-\bq)
    \,, \label{Eq:Blm}
}
where $Y^{TB}_{\l m,ij}(\hk)$ is the tensor spherical harmonics defined in Eq.~\eqref{Eq:YTB_ij} and
\al{
    T_\l^B(k) = \frac{6\sqrt{2}(-\iu)^{\l+1}}{\rho_{\gamma,0}}R_\gamma\left(\ln\frac{\eta_\nu}{\eta_B}\right)k^2\Delta_\l^{B,T}(k)
    \,.
}
In this paper, we assume that $1/\eta_B$ corresponds to the energy scale of grand unification and $\eta_\nu/\eta_B=10^{17}$ \cite{Shiraishi:2012rm,Shiraishi:2012:helical-PMF}

\subsection{Faraday rotation}

Faraday rotation to the CMB polarization at recombination by PMF is given by \cite{Kosowsky:2004zh}
\al{
    \alpha(\hatn) &= \frac{3}{16\pi^2\nu_0^2q}\INT{}{\eta}{}{\eta_*}{\eta_0}\dot{\tau}(\eta)\MB^i((\eta_0-\eta)\hatn)\hatn_i
    \,, 
}
where $q$ is the electron charge, $\nu_0$ is the observed frequency of CMB, $\dot{\tau}=an_{\rm e}\sigma_{\rm T}$ with $n_{\rm e}$ and $\sigma_{\rm T}$ being the free electron number density and the Thomson scattering cross section, respectively, and $\eta_*$ and $\eta_0$ are the conformal time at the recombination and at present, respectively. Following Ref.~\cite{Kosowsky:2004zh}, approximating that the contribution from the Faraday rotation comes from the PMFs at the recombination epoch. Then, the harmonic coefficients of the rotation angle becomes \cite{Kosowsky:2004zh}
\al{
    \alpha_{\l m} 
    &= \Int{}{k}{} T^\alpha_\l(k) \Int{2}{\hk}{} Y_{\l m,i}^{E,*}(\hk)\MB^i(\bk)
    \,, \label{Eq:alm}
}
where $Y^E_{\l m,i}(\hk)$ is the vector spherical harmonics defined in Eq.~\eqref{Eq:vec:Y^E} and we define
\al{
    T^\alpha_\l(k) \equiv -(-\iu)^{\l+1}k^2\frac{3}{32\pi^4\nu_0^2q}\sqrt{\l(\l+1)}\frac{j_\l(k\eta_0)}{k\eta_0}
    \,. 
}

\section{Cross bispectrum between CMB B-mode and Faraday rotation}
\label{sec:bispec}

We next consider the cross-bispectrum between B-mode and FR defined as
\al{
    b^{\l\l_1\l_2}_{mm_1m_2} = \ave{B_{\l m}\alpha_{\l_1m_1}\alpha_{\l_2m_2}}
    \,, \label{Eq:full-bispec:def}
}
where $\ave{\cdots}$ denotes the ensemble average. 
This correlation originates from the magnetic fields of the tensor passive mode. In this section, we derive the reduced bispectrum
\al{
    b_{\l\l_1\l_2} = \sum_{m_1,m_2,m_3}\Wjm{\l}{\l_1}{\l_2}{m}{m_1}{m_2}b^{\l\l_1\l_2}_{mm_1m_2}
    \,. \label{Eq:reduced-bispec:def}
}

Substituting Eqs.~\eqref{Eq:Blm} and \eqref{Eq:alm} into \eqref{Eq:full-bispec:def}, the B-mode FR cross-bispectrum is given by
\al{
    &b^{\l\l_1\l_2}_{mm_1m_2} 
    = \Int{}{k}{}T^B_\l(k)\Int{}{k_1}{} T^\alpha_{\l_1}(k_1) \Int{}{k_2}{} T^\alpha_{\l_2}(k_2) 
    \notag \\
    &\qquad\times\Int{2}{\hk{\rm d}^2\hk_1{\rm d}^2\hk_2}{}Y^{TB,*}_{\l m,ij}(\hk)Y_{\l_1m_1,a}^{E,*}(\hk_1)Y_{\l_2m_2,b}^{E,*}(\hk_2)
    \notag \\
    &\qquad\times\Int{3}{\bm{q}}{(2\pi)^6}\ave{\MB^i(\bm{q})\MB^j(\bk-\bm{q})\MB^a(\bk_1)\MB^b(\bk_2)}
    \,. \label{Eq:bispec:Baa:base}
}
Following usual conventions, we characterize the statistics of the PMFs using the following power spectrum of the magnetic fields \cite{Shaw2010MassiveUniverse,Planck:2015zrl}:
\al{
    \ave{\MB_a(\bk)\MB_b(\bk')} 
    &= \frac{1}{2}(2\pi)^3\delta(\bk+\bk')
    \notag \\
    &\quad\times\left[p_{ab,\hk}P_{\MB}(k)+\iu\epsilon_{abc}\hat{k}_cP_{\MH}(k)\right]
    \,, \label{Eq:P_B,P_H:def}
}
where we define $p_{ab,\hk}=\delta_{ab}-\hk_a\hk_b$, $\epsilon_{abc}$ is the Levi-Civita tensor, and $P_{\MB}(k)$ and $P_{\MH}(k)$ are the power spectrum of the non-helical and helical magnetic fields, respectively. We assume that the PMFs are Gaussian random fields. Using Wick's theorem and substituting Eq.~\eqref{Eq:P_B,P_H:def} into Eq.~\eqref{Eq:bispec:Baa:base}, the four-point correlation function in \eqref{Eq:bispec:Baa:base} is decomposed into 
\al{
    &\Int{3}{\bm{q}}{(2\pi)^6}\ave{B^i(\bm{q})B^j(\bk-\bm{q})B^a(\bk_1)B^b(\bk_2)}
    \notag \\
    &= \frac{1}{4}\delta(\bk_1+\bk_2+\bk)
    \notag \\
    &\qquad\times \bigg\{\left(p^{ai}_{\hk_1}p^{bj}_{\hk_2}+p^{aj}_{\hk_1}p^{bi}_{\hk_2}\right)P_{\MB}(k_1)P_{\MB}(k_2) 
    \notag \\
    &\qquad+ \left(q^{ai}_{\hk_1}q^{bj}_{\hk_2}+q^{aj}_{\hk_1}q^{bi}_{\hk_2}\right)P_{\MH}(k_1)P_{\MH}(k_2)
    \notag \\
    &\qquad+\left[p^{ai}_{\hk_1}q^{bj}_{\hk_2}+(i\leftrightarrow j)\right]P_{\MB}(k_1)P_{\MH}(k_2)
    \notag \\
    &\qquad+\left[p^{bj}_{\hk_2}q^{ai}_{\hk_1}+(i\leftrightarrow j)\right]P_{\MB}(k_2)P_{\MH}(k_1)
    \bigg\}
    \,, \label{Eq:4-point-B}
}
where $q^{ab}_{\hk}=\iu\epsilon^{abc}\hk_c$. The bispectrum is decomposed into the terms involving $P_{\MH}P_{\MH}$, $P_{\MH}P_{\MB}$, and $P_{\MB}P_{\MB}$ (hereafter, we call them the HH, BH and BB terms, respectively). We derive them separately as follows.

\subsection{HH term}

We first focus on the HH term that involves $P_{\MH}P_{\MH}$. In Eq.~\eqref{Eq:4-point-B}, the term involving $P_{\MH}P_{\MH}$ contains $q^{ai}_{\hk_1}q^{bj}_{\hk_2}$. Using 
\al{
    q^{ai}_{\hk}Y_{\l m,a}^{E,*}(\hk) &= Y_{\l m}^{B,i,*}(\hk)
    \,, \label{Eq:qxY^E}
}
the $E$-type vector spherical harmonics in Eq.~\eqref{Eq:bispec:Baa:base} become the $B$-type vector spherical harmonics. Since the delta function is the Fourier transform of the plane wave
\al{
    \delta(\bk+\bk_1+\bk_2) = \Int{3}{\br}{(2\pi)^3}\E^{-\iu\br\cdot(\bk+\bk_1+\bk_2)}
    \,, 
}
we obtain
\al{
    &(\text{HH term}) 
    = 2\Int{}{k}{} T^B_\l(k)
    \notag \\
    &\quad\times 
    \Int{}{k_1}{} T^\alpha_{\l_1}(k_1)P_{\MH}(k_1) \Int{}{k_2}{}  T^\alpha_{\l_2}(k_2)P_{\MH}(k_2) \Int{3}{\br}{}
    \notag \\
    &\quad\times I^{TB}_{\l m,ij}(k,\br)
    \left[I^{B,i}_{\l_1m_1}(k_1,\br)I^{B,j}_{\l_2m_2}(k_2,\br)+(i\leftrightarrow j)\right]
    \,. \label{Eq:bispec:HH:base}
}
where we have defined
\al{
    I^{TB}_{\l m,ij}(k,\br) 
    &\equiv \frac{1}{4\pi}\Int{2}{\hk}{} Y^{TB,*}_{\l m,ij}(\hk) \E^{-\iu\br\cdot\bk}
    \,, \label{Eq:I^B_lm_ij:def}
}
and
\al{
    I^{X}_{\l m,i}(k,\br) &\equiv \frac{1}{4\pi}\Int{2}{\hk}{}Y_{\l m,i}^{X,*}(\hk)\E^{-\iu\br\cdot\bk} 
    \,,
}
where $X=L,E,B$. The above integrals can be simplified to (see Appendix \ref{app:harmonics})
\al{
    I^{TB}_{\l m,ij}(k,\br) 
    &= c^{TB}_\l(kr)Y^{TB,*}_{\l m,ij}(\hr)+c^{VB}_\l(kr)Y^{VB,*}_{\l m,ij}(\hr)
    \,, \label{Eq:I^B_lm_ij}
}
where the coefficients, $c^{TB}_\l$ and $c^{VB}_\l$, are defined using the spherical Bessel function $j_\l$ and its derivative $j'_\l$ as (Eq.94, \cite{Dai:2012bc})
\al{
    c^{TB}_\l(x) &\equiv (-\iu)^{\l-1}\left(\frac{1}{2} j'_\l(x)+\frac{1}{x}j_\l(x)\right)
    \,, \\
    c^{VB}_\l(x) &\equiv (-\iu)^{\l-1}\sqrt{(\l-1)(\l+2)}\frac{j_\l(x)}{x}
    \,.
}
We also note that \cite{QTAM}
\al{
    I^{B}_{\l m,i}(k,\br) &= c^B_\l(kr)Y_{\l m,i}^{B,*}(\hr) 
    \,, \label{Eq:J^B_lm_i} \\
    I^{E}_{\l m,i}(k,\br)  
    &= c^L_\l(kr) Y^{L,*}_{\l m,i}(\hr)+c^E_\l(kr) Y^{E,*}_{\l m,i}(\hr)
    \,. \label{Eq:J^E_lm_i:def}
}
with 
\al{
    c^L_\l(x) &\equiv -(-\iu)^{\l+1}\frac{\sqrt{\l(\l+1)}}{x}j_\l(x)
    \,, \\
    c^E_\l(x) &\equiv -(-\iu)^{\l+1}\left(\frac{j_\l(x)}{x}+j'_\l(x)\right) 
    \,, \\
    c^B_\l(x) &\equiv (-\iu)^\l j_\l(x) 
    \,. 
}
Substituting Eqs.~\eqref{Eq:I^B_lm_ij} and \eqref{Eq:J^B_lm_i} into Eq.~\eqref{Eq:bispec:HH:base}, we obtain
\al{
    &(\text{HH term}) 
    = 2\Int{}{k}{} T^B_\l(k)
    \notag \\
    &\qquad\times 
    \Int{}{k_1}{} T^\alpha_{\l_1}(k_1)P_{\MH}(k_1) \Int{}{k_2}{}  T^\alpha_{\l_2}(k_2)P_{\MH}(k_2)
    \notag \\
    &\qquad\times 
    \Int{3}{\br}{}c^{TB}_\l(kr)Y^{TB,*}_{\l m,ij}(\hr)
    \notag \\
    &\qquad\times 
    \left[c_{\l_1}^B(k_1r)Y^{B,i,*}_{\l_1m_1}(\hr)c_{\l_2}^B(k_2r)Y^{B,j,*}_{\l_2m_2}(\hr)+(i\leftrightarrow j)\right]
    \,. 
}
In the above, we remove the term involving $Y_{\l m,ij}^{VB}$ which is orthogonal to the vector spherical harmonics, $Y_{\l m,i}^B$. 
The above quantity is symmetric in terms of the exchange between $i$ and $j$. 

In the following, we frequently use
\al{
    \Upsilon^{TB}_\l(r) &\equiv r^2\Int{}{k}{} c^{TB}_\l(kr) T^B_\l(k)
    \,, \label{Eq:U^TB:def} \\
    \Upsilon^{VB}_\l(r) &\equiv r^2\Int{}{k}{} c^{VB}_\l(kr) T^B_\l(k)
    \,, \label{Eq:U^VB:def}
}
Similarly, we define
\al{
    \Xi^{Z,X}_\l(r) \equiv \Int{}{k}{} c^Z_{\l}(kr)T^\alpha_{\l}(k)P_X(k) 
    \,, \label{Eq:Xi:def}
}
where $Z=L,E,B$, $X=\MB,\MH$ 

Using $\Upsilon^{TB}$ in Eq.~\eqref{Eq:U^TB:def} and $\Xi^{B,\MH}$ in Eq.~\eqref{Eq:Xi:def}, we rewrite the above equation as 
\al{
    &(\text{HH term}) 
    = 4\Int{}{r}{}\Upsilon^{TB}_\l(r)\Xi^{B,\MH}_{\l_1}(r)\Xi^{B,\MH}_{\l_2}(r)
    \notag \\
    &\qquad\times \Int{2}{\hr}{}Y^{TB,*}_{\l m,ij}(\hr)Y^{B,i,*}_{\l_1m_1}(\hr)Y^{B,j,*}_{\l_2m_2}(\hr)
    \,. \label{Eq:bispec:Baa:PHPH}
}
Using Eq.~\eqref{Eq:G:TB,B,B}, we integrate the products of the tensor and vector spherical harmonics with respect to $\hr$. Using the relationship between the bispectrum and reduced bispectrum \eqref{Eq:reduced-bispec:def}, we then obtain the reduced bispectrum from the HH term as
\al{
    b^{\MH\MH}_{\l\l_1\l_2} 
    &= -2\iu p^-_{\l\l_1\l_2}\gamma_{\l\l_1\l_2}\Wjm{\l}{\l_1}{\l_2}{2}{-1}{-1}
    \notag \\
    &\qquad\times
    \Int{}{r}{}\Upsilon^{TB}_\l(r)\Xi^{B,\MH}_{\l_1}(r)\Xi^{B,\MH}_{\l_2}(r)
    \,,
}
where $p^-_{\l\l_1\l_2}$ is the parity-odd coefficient that becomes unity if $\l+\l_1+\l_2$ is odd and is zero otherwise, and $\gamma_{\l\l_1\l_2}=[(2\l+1)(2\l_1+1)(2\l_2+1)/4\pi]^{1/2}$. 

\subsection{BH term}

Next, we focus on the term containing $P_{\MB}P_{\MH}$. In Eq.~\eqref{Eq:4-point-B}, the term involving $P_{\MB}P_{\MH}$ contains, e.g., $p^{ai}_{\hk_1}q^{bj}_{\hk_2}$. Using Eq.~\eqref{Eq:qxY^E} and
\al{
    p^{ai}_{\hk}Y_{\l m,a}^{E,*}(\hk) &= Y_{\l m}^{E,i,*}(\hk)
    \,, \label{Eq:pxY^E}
}
we obtain
\al{
    &(\text{BH term}) 
    = 4\Int{}{k}{} T^B_\l(k)
    \notag \\
    &\quad\times
    \bigg\{\Int{}{k_1}{} T^\alpha_{\l_1}(k_1)P_{\MB}(k_1) \Int{}{k_2}{} T^\alpha_{\l_2}(k_2)P_{\MH}(k_2) \Int{3}{\br}{}
    \notag \\
    &\quad\times
    I^{TB}_{\l m,ij}(k,\br)I^{E,i}_{\l_1m_1}(k_1,\br)I^{B,j}_{\l_2m_2}(k_2,\br) + (1\leftrightarrow 2)\bigg\}
    \,. \label{Eq:bispec:Baa:PBPH}
}
Using Eqs.~\eqref{Eq:I^B_lm_ij}, \eqref{Eq:J^B_lm_i}, \eqref{Eq:J^E_lm_i:def}, and the orthogonality between $Y^{TB}$ and $Y^L$, and that between $Y^{VB}$ and $Y^E$, the product $I^{TB}_{\l m,ij}(k,\br)I^{E,i}_{\l_1m_1}(k_1,\br)I^{B,j}_{\l_2m_2}(k_2,\br)$ is expressed with the tensor and vector spherical harmonics. Using $\Upsilon^{TB}$ in Eq.~\eqref{Eq:U^TB:def}, $\Upsilon^{VB}$ in Eq.~\eqref{Eq:U^VB:def}, and $\Xi^{Z,X}$ in Eq.~\eqref{Eq:Xi:def}, we rewrite Eq.~\eqref{Eq:bispec:Baa:PBPH} as
\al{
    &(\text{BH term})  
    = 4\Int{}{r}{}\Upsilon^{TB}_\l(r)\bigg\{\Xi^{E,\MB}_{\l_1}(r)\Xi^{B,\MH}_{\l_2}(r)
    \notag \\
    &\qquad\times \Int{2}{\hr}{}Y^{TB,*}_{\l m,ij}(\hr)Y^{E,i*}_{\l_1m_1}(\hr)Y^{B,j*}_{\l_2m_2}(\hr) + (1\leftrightarrow 2)\bigg\}
    \notag \\
    &\quad+ 4\Int{}{r}{}\Upsilon^{VB}_\l(r)\bigg\{\Xi^{L,B}_{\l_1}(r)\Xi^{B,H}_{\l_2}(r)
    \notag \\
    &\qquad\times \Int{2}{\hr}{}Y^{VB,*}_{\l m,ij}(\hr)Y^{L,i,*}_{\l_1m_1}(\hr)Y^{B,j,*}_{\l_2m_2}(\hr) + (1\leftrightarrow 2)\bigg\}
    \,. 
}
Using Eqs.~\eqref{Eq:G:TB,E,B} and \eqref{Eq:G:VB,L,B}, we obtain the reduced bispectrum from the BH term as
\al{
    &b^{\MB\MH}_{\l\l_1\l_2} 
    = -2\iu p^+_{\l\l_1\l_2}\gamma_{\l\l_1\l_2}\bigg\{\bigg[\Wjm{\l}{\l_1}{\l_2}{2}{-1}{-1}
    \notag \\
    &\qquad\times
    \Int{}{r}{}\Upsilon^{TB}_\l(r)\Xi^{E,\MB}_{\l_1}(r)\Xi^{B,\MH}_{\l_2}(r) + (1\leftrightarrow 2)\bigg]
    \notag \\
    &\quad- \sqrt{2}\bigg[\Wjm{\l}{\l_1}{\l_2}{1}{0}{-1}
    \notag \\
    &\qquad\times
    \Int{}{r}{}\Upsilon^{VB}_\l(r)\Xi^{L,\MB}_{\l_1}(r)\Xi^{B,\MH}_{\l_2}(r) + (1\leftrightarrow 2)\bigg]\bigg\}
    \,. 
}

\subsection{BB term}

Finally, we focus on the term containing $P_{\MB}P_{\MB}$. Using Eq.~\eqref{Eq:pxY^E}, we find
\al{
    &(\text{BB term})  
    = 4\Int{}{k}{}T^B_\l(k)
    \notag \\
    &\quad\times
    \bigg\{\Int{}{k_1}{} T^\alpha_{\l_1}(k_1)P_{\MB}(k_1) \Int{}{k_2}{} T^\alpha_{\l_2}(k_2)P_{\MB}(k_2) \Int{3}{\br}{}
    \notag \\
    &\quad\times
    I^{TB}_{ij,\l m}(k,\br)I^{E,i}_{\l_1m_1}(k_1,\br)I^{E,j}_{\l_2m_2}(k_2,\br) + (1\leftrightarrow 2)\bigg\}
    \,. \label{Eq:bispec:Baa:PBPB}
}
Using Eqs.~\eqref{Eq:I^B_lm_ij} and \eqref{Eq:J^E_lm_i:def} to expand $I_{\l m,ij}$ and $J_{\l m}^{E,i}$ with the tensor and vector spherical harmonics, and then using the quantities defined in Eqs.~\eqref{Eq:U^TB:def}, \eqref{Eq:U^VB:def}, and \eqref{Eq:Xi:def}, the bispectrum from the BB term becomes
\al{
    &(\text{BB term}) 
    =4\Int{}{r}{}\Upsilon^{TB}_\l(r)\Xi^{E,\MB}_{\l_1}(r)\Xi^{E,\MB}_{\l_2}(r)
    \notag \\
    &\qquad\times \Int{2}{\hr}{}Y^{TB,*}_{\l m,ij}(\hr)Y^{E,i*}_{\l_1m_1}(\hr)Y^{E,j*}_{\l_2m_2}(\hr)
    \notag \\
    &\quad+ 4\Int{}{r}{}\Upsilon^{VB}_\l(r)\bigg\{\Xi^{L,B}_{\l_1}(r)\Xi^{E,B}_{\l_2}(r)
    \notag \\
    &\qquad\times
    \Int{2}{\hr}{}Y^{VB,*}_{\l m,ij}(\hr)Y^{L,i,*}_{\l_1m_1}(\hr)Y^{E,j,*}_{\l_2m_2}(\hr) + (1\leftrightarrow 2)\bigg\}
    \,. 
}
Using Eqs.~\eqref{Eq:G:TB,E,E} and \eqref{Eq:G:VB,L,E} to integrate the product of the tensor and vector spherical harmonics, the reduced bispectrum from the BB term becomes
\al{
    &b^{\MB\MB}_{\l\l_1\l_2} 
    = -2\iu p^-_{\l\l_1\l_2}\gamma_{\l\l_1\l_2}
    \notag \\
    &\quad\times\bigg\{\Wjm{\l}{\l_1}{\l_2}{2}{-1}{-1}
    \Int{}{r}{}\Upsilon^{TB}_\l(r)\Xi^{E,\MB}_{\l_1}(r)\Xi^{E,\MB}_{\l_2}(r)
    \notag \\
    &\quad -\sqrt{2}\bigg[\Wjm{\l}{\l_1}{\l_2}{1}{0}{-1}\Int{}{r}{}\Upsilon^{VB}_\l(r)\Xi^{L,\MB}_{\l_1}(r)\Xi^{E,\MB}_{\l_2}(r) \notag \\
    &\quad - (1\leftrightarrow 2)\bigg]\bigg\}
    \,. 
}

\subsection{Final bispectrum expression}

We finally combine the above three terms. It is convenient to decompose the bispectrum into the parity-odd and parity-even contributions. Thus, we write the reduced bispectrum for the even and odd parts as
\al{
    &b^{\rm even}_{\l\l_1\l_2} 
    = 2\gamma_{\l\l_1\l_2}A_BA_\alpha^2
    \notag \\
    &\quad\times
    \bigg\{-\left[\Wjm{\l}{\l_1}{\l_2}{2}{-1}{-1}\Int{}{r}{}\bar{\Upsilon}^{TB}_\l(r)\bar{\Xi}^{E,\MB}_{\l_1}(r)\bar{\Xi}^{B,\MH}_{\l_2}(r) + (1\leftrightarrow 2)\right]
    \notag \\
    &\quad+ \left[\Wjm{\l}{\l_1}{\l_2}{1}{0}{-1}\Int{}{r}{}\bar{\Upsilon}^{VB}_\l(r)\bar{\Xi}^{L,\MB}_{\l_1}(r)\bar{\Xi}^{B,\MH}_{\l_2}(r) + (1\leftrightarrow 2)\right]\bigg\}
    \,, 
}
and
\al{
    &b^{\rm odd}_{\l\l_1\l_2} 
    = -2\iu A_BA_\alpha^2\gamma_{\l\l_1\l_2}
    \notag \\
    &\quad\times\bigg\{\Wjm{\l}{\l_1}{\l_2}{2}{-1}{-1}\Int{}{r}{}\bar{\Upsilon}^{TB}_\l(r)
    \notag \\
    &\qquad\qquad\times\left[\bar{\Xi}^{B,\MH}_{\l_1}(r)\bar{\Xi}^{B,\MH}_{\l_2}(r)-\bar{\Xi}^{E,\MB}_{\l_1}(r)\bar{\Xi}^{E,\MB}_{\l_2}(r)\right]
    \notag \\
    &\quad+ \left[\Wjm{\l}{\l_1}{\l_2}{1}{0}{-1}\Int{}{r}{}\bar{\Upsilon}^{VB}_\l(r)\bar{\Xi}^{L,\MB}_{\l_1}(r)\bar{\Xi}^{E,\MB}_{\l_2}(r) - (1\leftrightarrow 2)\right]\bigg\}
    \,. 
}
Here, we define
\al{
    A_B &\equiv \frac{6\sqrt{2}}{\rho_{\gamma,0}}R_\gamma\ln\frac{\eta_\nu}{\eta_B} 
    \,, \\
    A_\alpha &\equiv \frac{3}{32\pi^4\nu_0^2q}
    \,,
}
and
\al{
    \bar{\Upsilon}^{TB}_\l(r) &= \frac{1}{2}r^2\Int{}{k}{} k^2 \left(j'_\l(kr)+2\frac{j_\l(kr)}{kr}\right) \Delta_\l^{B,T}(k)
    \,, \\
    \bar{\Upsilon}^{VB}_\l(r) &= \sqrt{2(\l-1)(\l+2)}\,r\Int{}{k}{} k j_\l(kr) \Delta_\l^{B,T}(k)
    \,, 
}
We also define
\al{
    &\bar{\Xi}^{E,X}_\l(r) = \sqrt{\l(\l+1)} 
    \notag \\
    &\qquad \times\Int{}{k}{} k^2\left(\frac{j_\l(kr)}{kr}+j'_\l(kr)\right)\frac{j_\l(k\eta_0)}{k\eta_0} P_X(k) 
    \,, \\
    &\bar{\Xi}^{B,X}_\l(r) = \sqrt{\l(\l+1)} \Int{}{k}{} k^2 j_\l(kr)\frac{j_\l(k\eta_0)}{k\eta_0} P_X(k) 
    \,, \\
    &\bar{\Xi}^{L,X}_\l(r) = \l(\l+1)\Int{}{k}{} k^2 \frac{j_\l(kr)}{kr} \frac{j_\l(k\eta_0)}{k\eta_0} P_X(k) 
    \,. 
}

\section{Forecast}
\label{sec:forecast}

\begin{figure*}[t]
\bc
\includegraphics[width=8cm,clip]{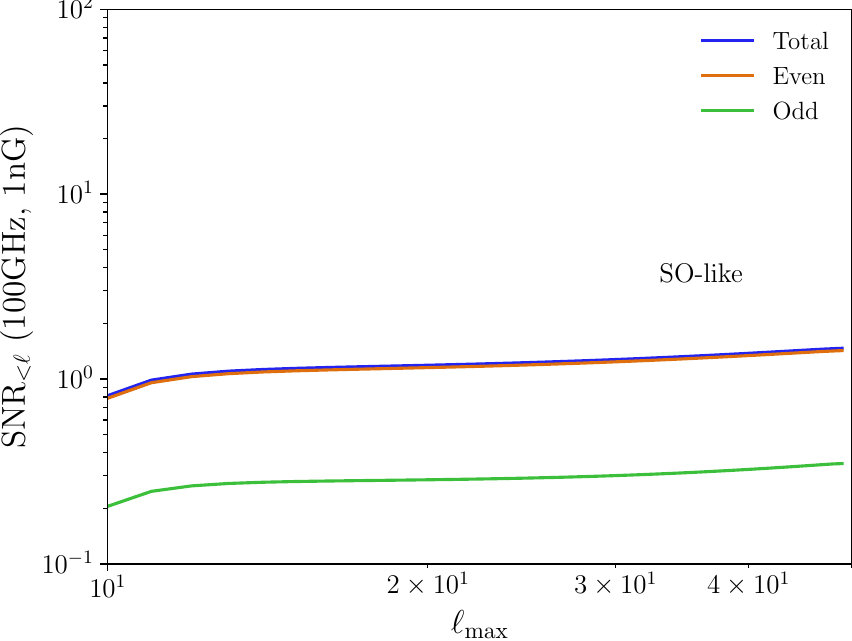}
\includegraphics[width=8cm,clip]{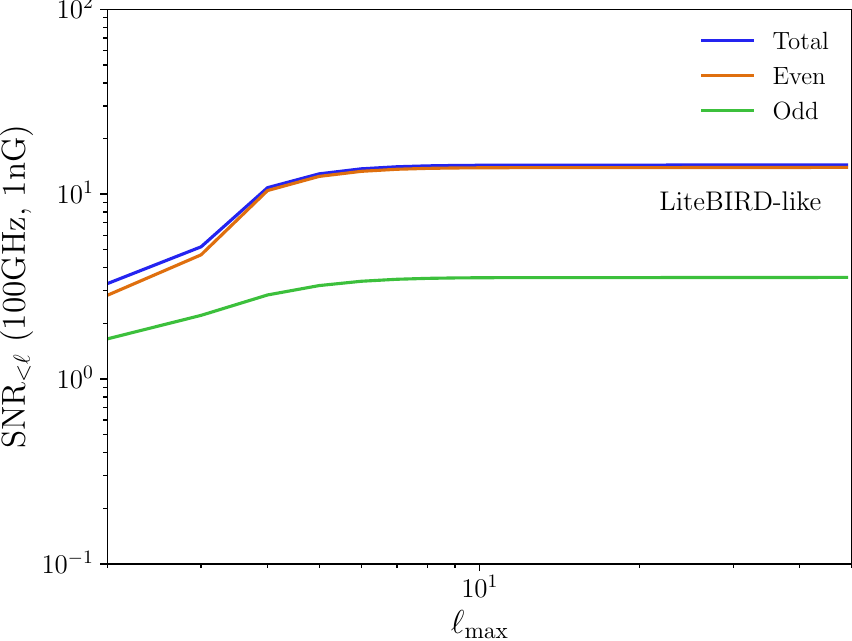}
\caption{SNR of the cross-bispectrum between the CMB $B$-mode polarization and FR angles as a function of the maximum multipole. We assume the SO- and LiteBIRD-like experiment and the FR angles are reconstructed at $100\,$GHz frequency band. The PMF amplitude is chosen as $1$nG. We also show the SNR for the parity-even and parity-odd contributions.}
\label{fig:snr}
\ec
\end{figure*}

We now discuss the sensitivity of the bispectrum to PMFs in ongoing and future CMB experiments. The signal-to-noise ratio (SNR) of the cross-bispectrum is defined as \cite{Hu:2000}
\al{
    \left(\frac{S}{N}\right)^2 = f_{\rm sky}\sum_{\l\l_1\l_2}\frac{b^2_{\l\l_1\l_2}}{2\hC_{\l}^{BB}\hC_{\l_1}^{\alpha\alpha}\hC_{\l_2}^{\alpha\alpha}}
    \,. 
}
where $f_{\rm sky}$ is the sky fraction of data used for the analysis, and $\hC^{BB}$ and $\hC^{\alpha\alpha}$ are the signal plus noise power spectra of the $B$ mode and FR, respectively. 
In our forecast, we include lensing signal in $\hC^{BB}$ and ignore Galactic foregrounds and inflationary $B$-modes. 
We also ignore PMF contributions to the cosmic variance. 

We assume that the power spectra of the non-helical and helical PMFs are given by \cite{Yura:2025}:
\al{
    P_{\MB}(k) &= A_{\MB}k^{n} 
    \,, \\
    P_{\MH}(k) &= r_HA_{\MB}k^{n} 
    \,, 
}
where the spectral index $n$ is assumed to be close to scale-invariant, $n=-2.9$. 
Note that $|r_H|\leq 1$ and maximal helicity is achieved if $|r_H|=1$. 
Using the smoothing scale $\lambda$, we reparametrize the amplitude $A_{\MB}$ as \cite{Caprini:2003vc,LiteBIRD:2024:PMF} 
\al{
    B^2_\lambda &= \INT{}{k}{}{0}{\infty} \frac{k^2}{2\pi^2} \E^{-k^2\lambda^2}P_{\MB}(k) = \frac{A_{\MB}}{4\pi^2\lambda^{n+3}}\Gamma\left(\frac{n+3}{2}\right)
    \,. 
}
such that $A_{\MB} = 4\pi^2\lambda^{n+3}\MB^2_\lambda/\Gamma\left(\frac{n+3}{2}\right)$. We consider the Simons Observatory (SO)-like and LiteBIRD-like experiments. 
For the SO-like experiment, we assume that $B$ modes for the bispectrum are measured from a small aperture telescope (SAT) with $1\mu$K-arcmin noise level 
in polarization with $30$ arcmin Gaussian beam \cite{SimonsObservatory:2025:SAT}, and that the minimum multipole of $B$ modes is $\l_{\rm min}=10$. The $B$-mode is further delensed and $70\%$ of the lensing $B$-mode is removed \cite{Namikawa:2021:SO-delens,SimonsObservatory:2024:SO-delens}. In the SO-like experiment, FR angle is reconstructed from a large aperature telescope (LAT) at $100$\,GHz. We assume $7.5\mu$K-arcmin polarization noise and $2$ arcmin Gaussian beam for the LAT $100\,$GHz data \cite{SimonsObservatory:2025:LAT}. We use {\tt cmblensplus} \cite{cmblensplus} to compute the noise power spectrum of the reconstructed FR angle $N_L^{\alpha\alpha}$ based on the $EB$ quadratic estimator \cite{Kamionkowski:2008fp,Namikawa:2016:rot,Kramer:2026:biref} where the CMB polarization multipoles at $500\leq\l\leq2000$ are used for the reconstruction which is similar to that adopted in the birefringence analysis of Ref.~\cite{Namikawa:2020:ACTDR4-biref}. We then assume $\l_{\rm min}=10$ for the reconstructed FR angles. We consider the overlapped region of SO-SAT and SO-LAT, $f_{\rm sky}=0.1$. 
For the LiteBIRD-like experiment, we assume that the white noise level in polarization is $2.5\mu$K-arcmin and the FWHM width of the Gaussian beam at $100\,$GHz is $24$ arcmin \cite{LiteBIRD:2024:PMF}. The $B$-mode is delensed and $60\%$ of the lensing $B$-mode is removed \cite{LiteBIRD:2023aov}. We assume $f_{\rm sky}=0.7$ for the LiteBIRD-like experiment. The CMB polarization at $2\leq\l\leq1000$ is used for reconstructing the FR angles with the $EB$ quadratic estimator. 

Figure~\ref{fig:snr} shows the cumulative SNR of the cross-bispectrum between the CMB $B$-mode polarization and the FR angle as a function of the maximum multipole, $\l_{\rm max}$, for the SO-like and LiteBIRD-like experiments. The FR angle is assumed to be reconstructed using the $100$ GHz frequency band, and the primordial magnetic-field amplitude is fixed at $B_{1{\rm Mpc}}=1$\,nG with the maximal helicity $r_H=1$. For the SO-like case, the total SNR increases gradually with $\l_{\rm max}$, reaching $\mC{O}(1)$ at the largest multipoles considered. The parity-even $P_\mathcal{B}P_\mathcal{H}$ contribution dominates the total signal, while the parity-odd $P_\mathcal{B}P_\mathcal{B}+P_\mathcal{H}P_\mathcal{H}$ contribution is smaller but remains non-negligible. In contrast, the LiteBIRD-like experiment yields a much larger SNR, with the total SNR exhibiting a sharp increase at low multipoles and reaching $\mC{O}(10)$. The parity-even component nearly saturates the total SNR, whereas the parity-odd component reaches only a few tens. These results indicate that the detectability of the cross-bispectrum is predominantly driven by the parity-even configurations, although the parity-odd sector can also provide an appreciable contribution. The smaller SNR of the parity-odd bispectrum compared with that of the parity-even bispectrum can be understood as follows. The parity-odd bispectrum is characterized by antisymmetric configurations and satisfies $b^{\rm odd}_{\l\l_1\l_2}=-b^{\rm odd}_{\l\l_2\l_1}$, i.e., $b^{\rm odd}_{\l\l_1\l_2}=0$ if $\l_1=\l_2$. 
For nearly symmetric configurations, $\ell_1 \simeq \ell_2$, 
the bispectrum is nearly symmetric and therefore contributes little to the parity-odd bispectrum. Consequently, the number of configurations that contribute significantly to the parity-odd bispectrum is reduced, resulting in a smaller SNR than that of the parity-even bispectrum. 

We note that the SNR is scaled with
\al{
    \left(\frac{S}{N}\right) \propto \left(\frac{B_{1\,\rm Mpc}}{1\,{\rm nG}}\right)^4\left(\frac{\nu_0}{100\,{\rm GHz}}\right)^{-4}
    \,. 
}
The results of the SNR calculation show that the cross-bispectrum has a sensitivity to PMFs whose amplitude is $B_{1\,\rm Mpc}\agt 0.7$\,nG.

\section{Summary}
\label{sec:summary}

We derived the $B\alpha\alpha$ bispectrum generated by the PMFs, allowing for both non-helical and helical components of the PMF power spectrum. Assuming Gaussian PMFs, the bispectrum is decomposed into contributions proportional to $P_{\MB}P_{\MB}$, $P_{\MB}P_{\MH}$, and $P_{\MH}P_{\MH}$. We derived analytic expressions for these contributions and showed that they contribute to different parity sectors. In particular, the mixed non-helical--helical contribution appears in the parity-even bispectrum, whereas the purely non-helical and purely helical contributions appear in the parity-odd bispectrum. This parity dependence provides a way to extract information on magnetic helicity that is inaccessible from the Faraday rotation power spectrum alone. We also estimated the detectability of the cross-bispectrum for SO-like, LiteBIRD-like, and cosmic-variance-limited CMB experiments. Our forecasts show that the $B\alpha\alpha$ bispectrum can provide sensitivity to sub-nG scale PMFs in future CMB observations. The cross-bispectrum therefore provides an independent and complementary probe of PMFs and offers a new avenue for testing the helicity of PMFs. 

We focused on the tensor passive mode as a source of PMFs, while the vector magnetic mode \cite{Shaw2010MassiveUniverse} would be another important source of the $B\alpha\alpha$ bispectrum at small angular scales. The cross-bispectrum between CMB temperature/$E$-mode and FR angles would also be useful for probing PMFs. In our forecast of SNR, we ignored the Galactic foregrounds for simplicity. The SNR could be reduced if we used a noise level after component separation. These calculations are left for our future work.


\begin{acknowledgments}
TN acknowledges support from JSPS KAKENHI Grant No. JP25K00996, and No. JP26H00405, and from JST EXPERT-J, Japan Grant No. JPMJEX2508. The Kavli IPMU is supported by World Premier International Research Center Initiative (WPI Initiative), MEXT, Japan. GL would like to acknowledge support from the Cambridge International Isaac Newton Scholarship. This work uses resources of the National Energy Research Scientific Computing Center (NERSC). We acknowledge the use of the following public software packages: {\tt CAMB} \cite{Lewis:1999:camb} and {\tt cmblensplus} \cite{cmblensplus}. The authors used ChatGPT (OpenAI) to assist with editing the paper and with the debugging of \href{https://github.com/toshiyan/pmf-bispec}{computer code} used in this study. All suggestions by ChatGPT were reviewed, tested, and modified as necessary by the authors. 
\end{acknowledgments}

\section*{Data availability}
The data that support the findings of this article are not publicly available. The data are available from the authors upon reasonable request.

\onecolumngrid

\appendix

\section{Useful formulas for vector and tensor spherical harmonics}
\label{app:harmonics}

\subsection{Spin-weighted spherical harmonics}

Denoting $\bn$ as the covariant derivative on the sphere, the covariant derivative of the spherical harmonics is related to the spin spherical harmonics as
\al{ 
	\bn Y_{\l m} = \sqrt{\l(\l+1)}\frac{\bm{\epsilon}_+ Y^{-1}_{\l m}-\bm{\epsilon}_-Y^1_{\l m}}{\sqrt{2}} 
    \,, 
} 
where we define the unit polarization vector as
\al{
    \bm{\epsilon}_{\pm} = \frac{\bm{e}_\theta\pm\bm{e}_\varphi}{\sqrt{2}}
    \,. 
}
The outer products of the radial unit vector and polarization vectors:
\al{
    \hatn\times\bm{\epsilon}_\pm = \mp\iu\bm{\epsilon}_\pm
    \,. 
}
We obtain
\al{ 
	\frac{\iu}{\sqrt{\l(\l+1)}}\hatn\times\bn Y_{\l m} = \frac{\bm{\epsilon}_+ Y^{-1}_{\l m}+\bm{\epsilon}_-Y^1_{\l m}}{\sqrt{2}} 
    \,. 
} 
The spin-rising and lowering operators: 
\al{
    \eth &= -\sqrt{2}\bm{\epsilon}_+\cdot\bn
    \,, \\
    \bar{\eth} &= -\sqrt{2}\bm{\epsilon}_-\cdot\bn
    \,, 
}
and
\al{
    (-\sqrt{2})^2\bm{\epsilon}_{+,i}\bm{\epsilon}_{+,j}\bn_i\bn_j &= \eth^2
    \,, \\
    (-\sqrt{2})^2\bm{\epsilon}_{-,i}\bm{\epsilon}_{-,j}\bn_i\bn_j &= \bar{\eth}^2
    \,. 
}
The spin-weighted spherical harmonics: 
\al{ 
	Y^s_{\l m} &= \left[\frac{(\l-s)!}{(\l+s)!}\right]^{1/2} \eth^s Y_{\l m} \qquad (0\leq s\leq \l) 
	\notag \\ 
	&= \left[\frac{(\l+s)!}{(\l-s)!}\right]^{1/2}(-1)^s\bar{\eth}^{-s} Y_{\l m} \qquad (-\l \leq s\leq 0) 
	\,,
}
and
\al{
    Y^2_{\l m} &= \left[\frac{(\l-2)!}{(\l+2)!}\right]^{1/2} \eth^2 Y_{\l m}
    \\ 
    Y^1_{\l m} &= \left[\frac{(\l-1)!}{(\l+1)!}\right]^{1/2} \eth Y_{\l m}
    \\ 
    Y^{-1}_{\l m} &= -\left[\frac{(\l-1)!}{(\l+1)!}\right]^{1/2} \bar{\eth} Y_{\l m}
    \\ 
    Y^{-2}_{\l m} &= \left[\frac{(\l-2)!}{(\l+2)!}\right]^{1/2} \bar{\eth}^2 Y_{\l m}
    \,. 
}
The integral of three spherical harmonics: 
\al{
	\Int{2}{\hatn}{} Y^{s_1}_{\l_1m_1}(\hatn)Y^{s_2}_{\l_2m_2}(\hatn)Y^{s_3}_{\l_3m_3}(\hatn) 
	= \gamma_{\l_1\l_2\l_3} 
	\Wjm{\l_1}{\l_2}{\l_3}{-s_1}{-s_2}{-s_3} \Wjm{\l_1}{\l_2}{\l_3}{m_1}{m_2}{m_3}
	\,. \label{Eq:Ylm-Wjm}
}
Here, we define
\al{
    \gamma_{\l_1\l_2\l_3} &= \sqrt{\frac{(2\l_1+1)(2\l_2+1)(2\l_3+1)}{4\pi}}
    \,.
}

\subsection{Vector spherical harmonics}

The vector spherical harmonics: 
\al{
    Y^E_{\l m,i}(\hatn) 
    &= \frac{1}{\sqrt{\l(\l+1)}}\bn_i Y_{\l m}(\hatn)
    = \frac{\bm{\epsilon}_{+,i} Y^{-1}_{\l m}-\bm{\epsilon}_{-,i}Y^1_{\l m}}{\sqrt{2}}
    \,, \label{Eq:vec:Y^E} \\
    Y^B_{\l m,i}(\hatn) &= \frac{-\iu}{\sqrt{\l(\l+1)}}(\hatn\times\bn)_i Y_{\l m}(\hatn) = -\frac{\bm{\epsilon}_{+,i} Y^{-1}_{\l m}+\bm{\epsilon}_{-,i}Y^1_{\l m}}{\sqrt{2}} 
    \,, \label{Eq:vec:Y^B} \\
    Y^L_{\l m,i}(\hatn) &= \hatn_i Y_{\l m}(\hatn)
    \,. \label{Eq:vec:Y^L}
}

\subsection{Tensor spherical harmonics}

We define the polarization tensor as $e_{ij}^{\pm2}(\hatn)=\sqrt{2}\epsilon^\pm_i(\hatn)\epsilon^\pm_j(\hatn)$ and then define the tensor spherical harmonics used in this paper as (see Eq.(91) of Ref.~\cite{Dai:2012bc} for all five tensor spherical harmonics, i.e., the longitudinal mode ($L$), two vector modes ($VE$, $VB$), and two tensor modes ($TE$, $TB$)) \footnote{The definition of $Y^{TB}$ here is a factor of two smaller than that of Ref.~\cite{Dai:2012bc}.}
\al{
    Y^{TE}_{\l m,ij}(\hatn) &= \frac{1}{2\sqrt{2}}\left[e_{ij}^{+2}(\hatn)Y_{\l m}^{-2}(\hatn)+e_{ij}^{-2}(\hatn)Y_{\l m}^{+2}(\hatn)\right] 
    \label{Eq:YTE_ij} \,, \\
    Y^{TB}_{\l m,ij}(\hatn) &= \frac{\iu}{2\sqrt{2}}\left[e_{ij}^{+2}(\hatn)Y_{\l m}^{-2}(\hatn)-e_{ij}^{-2}(\hatn)Y_{\l m}^{+2}(\hatn)\right]
    \label{Eq:YTB_ij} \,, \\
    Y^{VB}_{\l m,ij}(\hatn) &= -\frac{\iu}{\sqrt{2}}\left[\hatn_iY^B_{\l m,j}(\hatn)+\hatn_jY^B_{\l m,i}(\hatn)\right]
    \label{Eq:YVB_ij} \,. 
}

In the cross-bispectrum calculation, we need to evaluate Eq.~\eqref{Eq:I^B_lm_ij:def}: 
\al{
    I^{TB}_{\l m,ij}(k,\br) 
    &\equiv \Int{2}{\hk}{} Y^{TB,*}_{\l m,ij}(\hk) \E^{-\iu\br\cdot\bk}
    \,. 
}
To perform the integration, we use the following equation \cite{Dai:2012bc}
\al{
    e^{\mp2}_{ij}(\hk)\E^{\iu\bk\cdot\br} = \sum_{\alpha=TE,TB}\sum_{\l m}4\pi\iu^\l \epsilon^{\mp2,ab}(\hk)Y^{\alpha,*}_{\l m,ab}(\hk)\Psi^{k,\alpha}_{\l m,ij}(\br)
    \,. \label{Eq:app:TAM-wave-expand}
}
Here, $\Psi^{k,\alpha}_{\l m,ij}(\br)$ are the TAM waves that are expressed in terms of the tensor spherical harmonics as shown in Eq.(94) of Ref.~\cite{Dai:2012bc}. In our calculation, we need the following equations: 
\al{
    \Psi^{k,TB}_{\l m,ij}(\br) = -\iu\sqrt{(\l-1)(\l+1)}\frac{j_\l(kr)}{kr}2Y^{VB}_{\l m,ij}(\hatn) - \iu\left(j'_\l(kr)+\frac{2}{kr}j_\l(kr)\right)Y^{TB}_{\l m,ij}(\hatn)
    \,. 
}
Multiplying the spin-2 spherical harmonics $Y^{\pm2}_{\l m}(\hk)$ to the both sides of Eq.~\eqref{Eq:app:TAM-wave-expand}, and then integrating over $\hk$, we obtain
\al{
    \Int{2}{\hk}{}\sqrt{2}[Y_{\l m,ij}^{TE}(\hk)\pm\iu Y_{\l m,ij}^{TB}(\hk)]\E^{\iu\br\cdot\bk} = 4\pi\iu^\l\frac{1}{\sqrt{2}}\left(\Psi^{k,TE}_{\l m,ij}(\br)\pm\iu\Psi^{k,TB}_{\l m,ij}(\br)\right)
    \,, 
}
where we use Eqs.~\eqref{Eq:YTE_ij} and \eqref{Eq:YTB_ij} to rewrite the left-hand side of the above equation in terms of the tensor spherical harmonics. On the right-hand side, we also use Eqs.~\eqref{Eq:YTE_ij} and \eqref{Eq:YTB_ij} to rewrite the tensor spherical harmonics in terms of the spin-2 spherical harmonics and adopt the orthogonality of the spin-2 spherical harmonics. Extracting the $TB$ contribution, we finally obtain Eq.~\eqref{Eq:I^B_lm_ij}.

\subsection{Angular multipole coupling}

We need to compute the following quantities:
\al{
    G^{\l\l_1\l_2,(TB,B,B)}_{mm_1m_2} &= \Int{2}{\hr}{}Y^{TB,*}_{\l m,ij}(\hr)Y^{B,i,*}_{\l_1m_1}(\hr)Y^{B,j,*}_{\l_2m_2}(\hr) 
    \,, \\
    G^{\l\l_1\l_2,(TB,E,B)}_{mm_1m_2} &= \Int{2}{\hr}{}Y^{TB,*}_{\l m,ij}(\hr)Y^{E,i,*}_{\l_1m_1}(\hr)Y^{B,j,*}_{\l_2m_2}(\hr) 
    \,, \\
    G^{\l\l_1\l_2,(VB,L,B)}_{mm_1m_2} &= \Int{2}{\hr}{}Y^{VB,*}_{\l m,ij}(\hr)Y^{L,i,*}_{\l_1m_1}(\hr)Y^{B,j,*}_{\l_2m_2}(\hr) 
    \,, \\
    G^{\l\l_1\l_2,(TB,E,E)}_{mm_1m_2} &= \Int{2}{\hr}{}Y^{TB,*}_{\l m,ij}(\hr)Y^{E,i,*}_{\l_1m_1}(\hr)Y^{E,j,*}_{\l_2m_2}(\hr) 
    \,, \\
    G^{\l\l_1\l_2,(VB,L,E)}_{mm_1m_2} &= \Int{2}{\hr}{}Y^{VB,*}_{\l m,ij}(\hr)Y^{L,i,*}_{\l_1m_1}(\hr)Y^{E,j,*}_{\l_2m_2}(\hr) 
    \,. 
}
Using Eqs.~\eqref{Eq:vec:Y^E}, \eqref{Eq:vec:Y^B}, \eqref{Eq:vec:Y^L}, \eqref{Eq:YTB_ij}, and \eqref{Eq:YVB_ij} to rewrite the vector and tensor spherical harmonics with the spin-weighted spherical harmonics, and then using Eq.~\eqref{Eq:Ylm-Wjm} to relate them with the Wigner $3j$ symbols, we obtain
\al{
    G^{\l\l_1\l_2,(TB,B,B)}_{mm_1m_2} &= -\frac{\iu}{2}p^-_{\l\l_1\l_2}\gamma_{\l\l_1\l_2} \Wjm{\l}{\l_1}{\l_2}{2}{-1}{-1}\Wjm{\l}{\l_1}{\l_2}{m}{m_1}{m_2} 
    \,, \label{Eq:G:TB,B,B} \\
    G^{\l\l_1\l_2,(TB,E,B)}_{mm_1m_2} &= -\frac{\iu}{2}p^+_{\l\l_1\l_2}\gamma_{\l\l_1\l_2} \Wjm{\l}{\l_1}{\l_2}{2}{-1}{-1}\Wjm{\l}{\l_1}{\l_2}{m}{m_1}{m_2} 
    \,, \label{Eq:G:TB,E,B} \\
    G^{\l\l_1\l_2,(VB,L,B)}_{mm_1m_2} &= \frac{\iu}{\sqrt{2}}p^+_{\l\l_1\l_2}\gamma_{\l\l_1\l_2} \Wjm{\l}{\l_1}{\l_2}{1}{0}{-1}\Wjm{\l}{\l_1}{\l_2}{m}{m_1}{m_2} 
    \,, \label{Eq:G:VB,L,B} \\
    G^{\l\l_1\l_2,(TB,E,E)}_{mm_1m_2} &= G^{\l\l_1\l_2,(TB,B,B)}_{mm_1m_2} 
    \,, \label{Eq:G:TB,E,E} \\
    G^{\l\l_1\l_2,(VB,L,E)}_{mm_1m_2} &= \frac{\iu}{\sqrt{2}}p^-_{\l\l_1\l_2}\gamma_{\l\l_1\l_2} \Wjm{\l}{\l_1}{\l_2}{1}{0}{-1}\Wjm{\l}{\l_1}{\l_2}{m}{m_1}{m_2} 
    \,. \label{Eq:G:VB,L,E} 
}

\bibliographystyle{mybst}
\bibliography{cite}

\end{document}